\documentclass[preprint, superscriptaddress, floatfix, preprintnumbers, nofootinbib, altaffilletter]{revtex4-1}
\usepackage{graphicx}
\usepackage{subfigure}
\usepackage{dcolumn}
\usepackage{bm}
\usepackage{amssymb}
\usepackage{amsmath}
\usepackage{xcolor}
\usepackage[utf8]{inputenc}
\usepackage[OT1]{fontenc}
\usepackage{yfonts,amsthm,amsfonts,amssymb,amscd, ulem}
\usepackage{enumerate}
\usepackage{fancyhdr}
\usepackage{mathtools}
\usepackage{mathrsfs}
\usepackage{cancel}
\usepackage{slashed}
\usepackage{bigints}
\usepackage[flushleft]{threeparttable}
\usepackage[colorlinks=true, citecolor=purple, linkcolor=blue]{hyperref}
\usepackage{url}
\usepackage{makecell,booktabs}
\usepackage{braket}
\usepackage{relsize}
\usepackage{multirow}
\usepackage{verbatim}
\usepackage{txfonts}
\usepackage{upgreek}
\usepackage{extarrows}
\usepackage{array}
\usepackage{appendix}
\usepackage[T1]{fontenc}
\usepackage{setspace}
\usepackage{float}
\usepackage{orcidlink}
\usepackage{tabularx}
\usepackage{array}
\usepackage{amsmath}

\renewcommand{\arraystretch}{0.6}

\begin{document}

\title{Probing anomalous $Wtb$ couplings through $bW$ spin correlations in top-quark pair production at the HL-LHC}

\author{Qi Bi}
\email{biqii@buaa.edu.cn}
\affiliation{School of Physics, Beihang University, Beijing 100191, China}

\author{Zekun Li}
\email{lizekun@stu.pku.edu.cn}
\affiliation{School of Physics and State Key Laboratory of Nuclear Physics and Technology, Peking University, Beijing 100871, China}

\author{Chuntao Liu}
\email{lchuntao@buaa.edu.cn}
\affiliation{School of Physics, Beihang University, Beijing 100191, China}

\author{Jia Liu \orcidlink{0000-0001-7386-0253}}
\email{jialiu@pku.edu.cn}
\affiliation{School of Physics and State Key Laboratory of Nuclear Physics and Technology, Peking University, Beijing 100871, China}
\affiliation{Center for High Energy Physics, Peking University, Beijing 100871, China}

\author{Xiao-Ping Wang \orcidlink{0000-0002-2258-7741}}
\email{hcwangxiaoping@buaa.edu.cn}
\affiliation{School of Physics, Beihang University, Beijing 100191, China}

\preprint{CPTNP-2026-024}

\begin{abstract}

We propose a spin-density-matrix framework for probing anomalous $Wtb$ interactions in top-quark pair production by exploiting the polarizations of the $b$ quark and the $W$ boson and their spin correlations. The $b$-quark spin is analyzed through angular distributions in semileptonic $\Lambda_b$ decay following fragmentation, while the leptonic $W$ decay serves as a $W$-polarization analyzer. We derive the $bW$ spin density matrix in the presence of anomalous $Wtb$ couplings and show that its polarization and spin-correlation parameters provide linear sensitivity to these couplings. We construct experimentally accessible angular observables associated with these parameters and evaluate their statistical sensitivities. For the HL-LHC at $\sqrt{s}=14~ \mathrm{TeV}$ with an integrated luminosity of $3000~ \mathrm{fb}^{-1}$, the proposed method yields improved constraints on most anomalous $W t b$ couplings, with the sensitivity to $V_R$ enhanced by a factor of 8 and that to $g_L$ by a factor of about 6.
\end{abstract}

\maketitle

\section{Introduction}
The Large Hadron Collider (LHC) has been instrumental in performing precision measurements of the Standard Model (SM) and searching for physics beyond the Standard Model (BSM). 
However, while the SM accounts for nearly all collider data with remarkable precision, it remains demonstrably incomplete as a definitive theory of nature.
Several fundamental questions, such as the origin of matter-antimatter asymmetry~\cite{Trodden:1998ym,Cohen:1993nk,Gavela:1993ts,Huet:1994jb,Gavela:1994ds,Gavela:1994dt,Riotto:1999yt}, neutrino masses~\cite{Davis:1968cp,KamLAND:2002uet,DayaBay:2012fng,DoubleChooz:2014kuw} and dark matter~\cite{Hooper:2009zm,Gorenstein:2014iba,Gelmini:2015zpa,Bertone:2004pz,Feng:2010gw,Arcadi:2017kky,Roszkowski:2017nbc}, remain unresolved. 
In the absence of direct evidence for new particles or interactions, indirect searches through precision measurements of SM processes offer a promising pathway to uncover subtle deviations that may signal new physics.

One particularly powerful approach involves studying the properties of the top quark, the heaviest known elementary particle, whose large mass and short lifetime make it a unique probe of the electroweak sector and potential new physics~\cite{Beneke:2000hk,Quadt:2006dqn,Bernreuther:2008ju}. 
Because the top quark decays before hadronization, its spin information is transferred directly to its decay products and can be extracted from their angular distributions, providing sensitive probes of the interaction structure~\cite{Jezabek:1994zv,Mahlon:1995zn,Mahlon:1996pn,Bernreuther:2010ny,Bernreuther:2015yna}.
In the top-quark sector, anomalous $Wtb$ interactions~\cite{delAguila:2002nf,Grzadkowski:2008mf,Aguilar-Saavedra:2006qvv,Aguilar-Saavedra:2008quj,Rindani:2011pk,PrasathV:2014omf,Cao:2015doa,Jueid:2018wnj,D0:2008zjk,CMS:2016uzc} can induce CP-violating effects in top-quark production and decay.
Precision measurements of the $Wtb$ vertex provide a window into exploring CP violation and help us to understand the mechanism for the matter-antimatter asymmetry.
The structure of the $Wtb$ vertex can be sensitively tested through measurements of top-quark polarization~\cite{Rindani:2015vya}, $W$-boson helicity fractions~\cite{Aguilar-Saavedra:2010ljg,Aguilar-Saavedra:2015yza}, and forward-backward asymmetries. Anomalous couplings in this vertex, parameterized within an effective field theory framework, can reveal deviations from SM expectations and offer critical insights into new physics at higher energy scales.

Recently, a parallel avenue of research has emerged at the intersection of high-energy physics and quantum information science. 
It has been shown that quantum entanglement~\cite{Einstein:1935rr,Schrodinger:2008pyl} and Bell nonlocality~\cite{Bell:1964kc} can be probed in high-energy colliders through spin correlations in heavy quark pairs~\cite{Afik:2020onf,Fabbrichesi:2021npl,Barr:2024djo,Severi:2021cnj,Afik:2022dgh,Afik:2022kwm,Han:2023fci,Aguilar-Saavedra:2022uye,Dong:2023xiw,Cheng:2024btk,Han:2024ugl,Fabbrichesi:2025psr,Afik:2026pxv}, $\tau$ lepton pairs ~\cite{Altakach:2022ywa,Ehataht:2023zzt,Ma:2023yvd,Fabbrichesi:2024wcd,Han:2025ewp,Morales:2024jhj,Zhang:2025mmm,Ai:2025wnt,Yang:2026uwu}, massive gauge boson pairs~\cite{Barr:2021zcp,Barr:2022wyq,Ashby-Pickering:2022umy,Aguilar-Saavedra:2022wam,Aguilar-Saavedra:2022mpg,Fabbrichesi:2023cev,Bernal:2023ruk,Bi:2023uop,Grossi:2024jae,Ding:2025mzj,Goncalves:2025mvl,Goncalves:2025xer},  baryons~\cite{Chen:2013epa,Fabbrichesi:2024rec,Wu:2024asu,Feng:2025ryr,Pei:2025yvr,Pei:2025ito,Hong:2025drg,Zhang:2026nwm} and light
quark pairs~\cite{Cheng:2025cuv,Hatta:2025obw}.
The ATLAS and CMS collaborations have already observed spin entanglement in top quark pairs~\cite{ATLAS:2023fsd,CMS:2024pts,CMS:2024zkc}.
Although quarks other than the top quark hadronize before decaying, information about their polarizations and spin correlations can still be experimentally accessed through hadronization products~\cite{Kats:2023zxb}.
In particular, $\Lambda_b$ baryons can retain a fraction of the polarization of their parent $b$ quarks, while their decay products serve as spin analyzers~\cite{Falk:1993rf,Manohar:1993qn,Galanti:2015pqa}.
The polarization retained by $\Lambda_b$ baryons allows $b$-quark polarization and $b\bar b$ spin correlations~\cite{Kats:2023zxb} to be extracted, providing a potential probe of entanglement and Bell nonlocality of $b$ quark pairs~\cite{Afik:2025grr}. In top-quark decay, combining $\Lambda_b$ and $W$ polarimetry can, in principle, determine all four $t\to Wb$ helicity amplitudes, including the relative phase between the left- and right-handed $b$ quark amplitudes~\cite{Nelson:2001su}.

In this work, we bridge these two frontiers by exploring how quantum information observables can be used to constrain anomalous couplings in the $Wtb$ interaction. 
We use the formalism developed for probing $b$ quark polarization and adapt it to the top quark sector, where the decay products directly reflect the spin state of the parent quark. By combining the sensitivity of quantum correlations with the precision of top quark property measurements, we aim to enhance the search for new physics in the high luminosity phase of the LHC and beyond.

We present a feasibility study of spin correlation measurements in the $bW^+$ and $\bar bW^-$ systems under the influence of anomalous $Wtb$ couplings at projected HL-LHC scenarios. Our results demonstrate that quantum information-based observables provide complementary information and often offer enhanced sensitivity to certain anomalous couplings, particularly the right-handed vector coupling $V_R$ and the left-handed tensor coupling $g_L$, which are poorly constrained by conventional observables alone.

This paper is organized as follows.
In Sec.\ref{sec:wtb}, we present the theoretical framework for the $Wtb$ interaction using a model-independent extension of the SM Lagrangian parameterized by the anomalous couplings $V_L$, $V_R$, $g_L$, and $g_R$.
We then introduce the spin density matrix description of the $bW$ system and derive its density matrix from top quark decay in both the SM and the anomalous $Wtb$ interaction.
In Sec.\ref{sec:matrix_to_observables}, we relate the components of the density matrix to measurable angular observables.
In Sec.\ref{sec:uncertainty}, we derive the corresponding statistical uncertainties for the observables.
In Sec.\ref{sec:analysis}, we describe the event analysis and estimate the effective number of signal events expected at the HL-LHC.
In Sec.\ref{sec:result}, we present the projected constraints on the anomalous couplings.
In Sec.\ref{sec:conclusion}, we conclude.

\section{Anomalous $Wtb$ interaction}
\label{sec:wtb}

The top quark decays almost exclusively through the charged-current process $t\rightarrow bW$, with a branching ratio close to unity. 
The $Wtb$ vertex therefore provides a unique probe of the electroweak interaction of the top quark and possible effects from physics beyond the Standard Model.
Due to its large mass and extremely short lifetime, the top quark decays before hadronization, allowing its spin information to be directly encoded in the angular distributions of the decay products. Consequently, deviations from the SM structure of the $Wtb$ interaction can manifest themselves through modifications of the polarizations and spin correlations of the final-state $bW$ system.

\subsection{Effective Lagrangian}
In this work, we adopt a model-independent effective-field-theory parametrization of the $Wtb$ vertex and investigate how anomalous $Wtb$ couplings affect the spin structure of the $bW$ system. The most general interaction involving dimension-four and dimension-five operators can be written as
\begin{equation}
\mathcal{L}_{\mathrm{Wtb}}=-\frac{g_W}{\sqrt{2}}\bar{b}\,\gamma^\mu (V_LP_L+V_RP_R)\,t\,W_\mu^--\frac{g_W}{\sqrt{2}}\bar{b}\,\frac{i\sigma^{\mu\nu}p_{\nu}}{M_W} (g_LP_L+g_RP_R)\,t\,W_\mu^-+\mathrm{h.c.},
\end{equation}
where $V_{L,R}$ and $g_{L,R}$ denote left- and right-handed vector and tensor couplings, respectively.
The SM interaction corresponds to the special case $V_L=1$ and $V_R=g_L=g_R=0$. With the general $Wtb$ interaction, the decay amplitude of the top quark takes the form
\begin{equation}
\mathcal{M}_{\beta\lambda,\alpha}=-\frac{g_W}{\sqrt{2}}\varepsilon^*_\mu(p_W,\lambda)\cdot \bar{u}_b(p_b,\beta)\left[\gamma^\mu(V_LP_L+V_RP_R)+\frac{i\sigma^{\mu\nu}p_{W,\nu}}{M_W} (g_LP_L+g_RP_R)\right]u_t(p_t,\alpha).
\end{equation}
Here, $p_t$, $p_b$, and $p_W$ denote the four-momenta of the top quark, bottom quark, and $W$ boson, respectively, with $p_W=p_t-p_b$. 
The indices $\alpha$, $\beta$, and $\lambda$ label the spin of the top quark, the spin of the bottom quark, and the polarization of the $W$ boson, respectively.

\subsection{Density Matrix of $bW$ System}\label{sec:density_matrix}

Since both the $b$ quark and the $W$ boson carry spin, the $bW$ final state forms a bipartite quantum system whose spin structure encodes detailed information about the underlying decay dynamics. 
A natural and model-independent description of these spin effects is provided by the density matrix formalism, in which the $bW$ system is treated as a composite state in the $2\times3$-dimensional Hilbert space corresponding to a spin-$\frac12$ fermion and a spin-1 vector boson. 

The most general density operator can be parameterized as
\begin{equation}
\rho^{\mathrm{bW}} = \frac{1}{6} I_2\otimes I_3 + \frac{1}{6}\tilde{B}_i\, \sigma_i \otimes I_3  + \frac{1}{2}\tilde{d}_i\, I_2\otimes S_i +\frac{1}{2}\tilde{q}_{ij}\,I_2\otimes S_{ij} + \frac{1}{2} \tilde{C}_{ij}\, \sigma_i\otimes S_j + \frac{1}{2} \tilde{C}_{jk}^i\,\sigma_i\otimes S_{jk} ,
\end{equation}
where $I_{2(3)}$ denotes the identity matrix in the spin- $\frac{1}{2}$ (spin-1) space, $\sigma_i$ are the Pauli matrices, and $S_i$ are the generators of the spin-1 representation\footnote{
For explicit calculations, it is convenient to work in the basis of eigenstates of $S_z$.
\[
\begin{aligned}
S_x &= \tfrac{1}{\sqrt{2}}\begin{pmatrix}
0&1&0\\
1&0&1\\
0&1&0
\end{pmatrix},\quad
S_y = \tfrac{1}{\sqrt{2}}\begin{pmatrix}
0&-i&0\\
i&0&-i\\
0&i&0
\end{pmatrix},\quad
S_z = \begin{pmatrix}
1&0&0\\
0&0&0\\
0&0&-1
\end{pmatrix}.
\end{aligned}
\]
}.
The rank-2 operators are defined as $S_{i j} \equiv S_i S_j+S_j S_i$, and $\tilde{q}_{i j}$ are symmetric and traceless.
Thus, there are $6\times 6-1=35$ independent parameters that can be extracted from the density matrix, each possessing a distinct physical interpretation. Specifically, $\tilde{B}_i$ describes the polarization of the $b$ quark; $\tilde{d}_i$ and $\tilde{q}_{i j}$ encode the polarization and quadrupole alignment of the $W$ boson, respectively; $\tilde{C}_{i j}$ and $\tilde{C}_{j k}^i$ characterize the spin correlations between the two particles.

\begin{figure}
    \centering
    \includegraphics[width=0.9\linewidth]{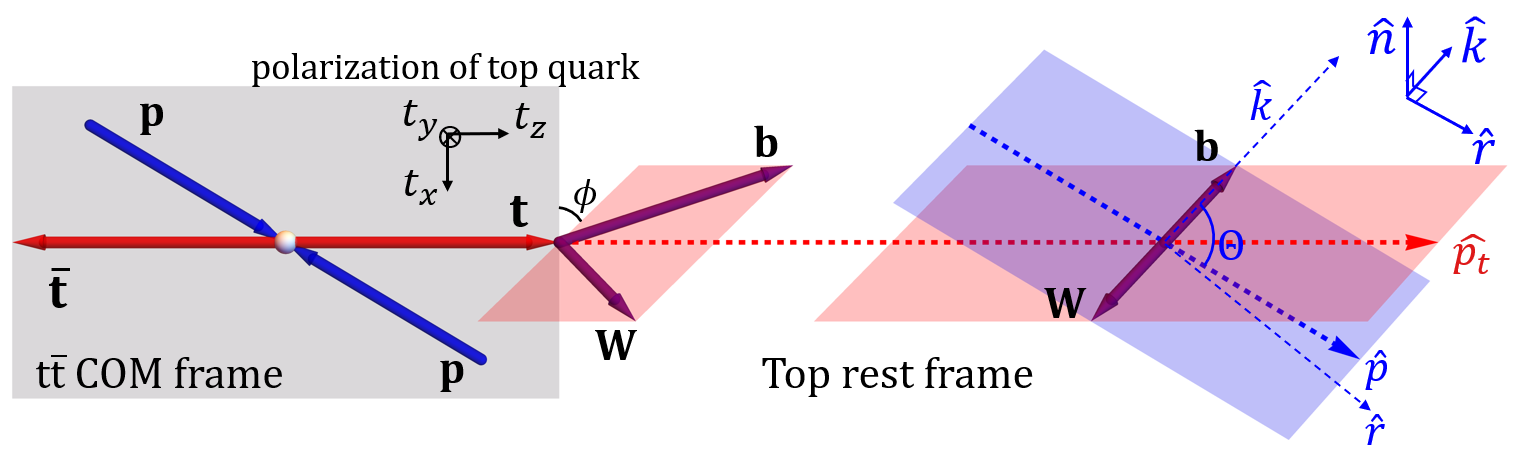}
    \caption{The schematic illustration of the top-quark pair production and decay process at colliders.
    The blue vectors denote the momenta of incoming partons in the $t\bar t$ COM frame while the red vectors denote the momenta of $t$ quark and $\bar{t}$ quark. The purple vectors represent the momenta of $b$ quark and $W$ boson. 
    And the basis of polarization vectors of $t$ is defined in the scattering plane\cite{ATLAS:2026vxj}.}
    \label{fig:schematic}
\end{figure}

The coefficients in the density matrix are defined in the top-quark rest frame. 
To construct measurable angular observables, we introduce a reference basis determined by the kinematics of the $bW$ system. 
The longitudinal axis is chosen along the momentum direction of the bottom quark,
\begin{equation}
\hat{k}=\frac{\mathbf{p}_b}{|\mathbf{p}_b|},
\end{equation}
while the transverse axes are defined with respect to the incoming proton
direction $\hat{p}$,
\begin{equation}
\hat{r}
=
\frac{\hat{p}-\hat{k}\cos\Theta}{\sin\Theta},
\qquad
\hat{n}=\hat{r}\times\hat{k},
\end{equation}
where $\cos\Theta=\hat{p}\cdot\hat{k}$. 
This choice defines a right-handed coordinate system $\{\hat n,\hat r,\hat k\}$ in the top-quark rest frame, which is used throughout the analysis and illustrated schematically in Fig.~\ref{fig:schematic}. 
This basis is analogous to the helicity basis commonly used in top-quark spin analyses, where the $b$-quark momentum defines the longitudinal direction. 
It provides a direct connection between the components of the $bW$ density matrix and the angular distributions of the $b$- and $W$-decay products.

Given a $2\times2$ initial top-quark density matrix $\rho^\mathrm t$, the $6\times 6$ density matrix of the final-state $bW$ system can be theoretically calculated by
\begin{equation}\label{eq.rho_general}
\rho^{\mathrm{bW}}_{\beta\lambda,\beta'\lambda'}=\mathcal{M}_{\beta\lambda,\alpha}\,\rho^\mathrm t_{\alpha\alpha'}\,\mathcal{M}^\dagger_{\alpha',\beta'\lambda'},
\end{equation}
where $\alpha,\alpha'=\pm1,\, \beta,\beta'=\pm1,\, \lambda,\lambda'=+1,0,-1$ are the helicity indices of top quark, bottom quark, and $W$ boson, respectively.
For an unpolarized top-quark state, the spin density matrix is given by $\rho^t=I_2/2$, where $I_2$ is the $2\times2$ identity matrix.

\subsection{Dependence on Anomalous Couplings}
We now explicitly investigate how the anomalous $Wtb$ couplings modify the spin density matrix $\rho^{\mathrm{bW}}$ of the $bW$ system. In order to show the physics more clearly, we choose the right-handed vector coupling $V_R$ to be real, with $V_L=1$ and $g_L=g_R=0$. The density matrix takes the form
\begin{align}
    \rho^{\mathrm{bW}}(V_R)= \frac{1}{N}
    \begin{pmatrix}
        0 & 0 & 0 & 0 & 0 & 0 \\
        0 & A & 0 & E & 0 & 0 \\
        0 & 0 & B & 0 & F & 0 \\
        0 & E & 0 & C & 0 & 0 \\
        0 & 0 & F & 0 & D & 0 \\
        0 & 0 & 0 & 0 & 0 & 0 \\
    \end{pmatrix},
\label{density-dependece}
\end{align}
where $N$ is the normalization factor.
The nonvanishing matrix elements (A)-(F), which describe the populations and interference terms of the allowed helicity configurations, are given by
\begin{equation}
\begin{aligned}
    &A = \frac{1}{2m_W^2}((m_t x + m_b y)^2 + 2V_R(m_t x + m_b y )(m_b x + m_t y) + V_R^2(m_b x + m_t y)^2),\\
    &B = x^2 + 2V_R xy+V_R^2 y^2, \\
    &C = y^2 + 2V_R xy+V_R^2 x^2,\\
    &D = \frac{1}{2m_W^2}((m_b x + m_t y)^2 + 2V_R(m_t x + m_b y )(m_b x + m_t y) + V_R^2(m_t x + m_b y)^2),\\
    &E = \frac{1}{\sqrt{2}m_W}(y(m_t x+m_b y) + V_R(m_t x^2 + 2m_b x y + m_t y^2)+ V_R^2 x (m_b x + m_t y)), \\
    &F = \frac{1}{\sqrt{2}m_W}(x(m_b x+m_t y) + V_R(m_t x^2 + 2m_b x y + m_t y^2)+ V_R^2 y (m_t x + m_b y)), \\
\end{aligned}
\end{equation}
with
\begin{equation}
\begin{aligned}
&x = m_t\left[ \sqrt{\left(1-\frac{m_b}{m_t}\right)^2-\left(\frac{m_W}{m_t}\right)^2}-\sqrt{\left(1+\frac{m_b}{m_t}\right)^2-\left(\frac{m_W}{m_t}\right)^2}\right],\\
&y = m_t\left[ \sqrt{\left(1-\frac{m_b}{m_t}\right)^2-\left(\frac{m_W}{m_t}\right)^2}+\sqrt{\left(1+\frac{m_b}{m_t}\right)^2-\left(\frac{m_W}{m_t}\right)^2}\right].
\end{aligned}
\end{equation}

For sufficiently small anomalous couplings, the dominant sensitivity typically arises from linear terms generated by interference between the SM and anomalous amplitudes.
Since these terms change sign when the signs of the anomalous couplings are reversed, they provide sensitivity to both the magnitude and the sign of the anomalous coupling.
In contrast, when the interference contribution is absent or strongly suppressed, the leading dependence becomes quadratic, resulting in reduced sensitivity to the coupling.

The dependence on $V_R$ provides a particularly instructive example. In the limit $m_b\rightarrow0$, the kinematic quantities defined above reduce to
\begin{align}
x\rightarrow0,
\qquad
y\rightarrow2\sqrt{m_t^2-m_W^2}.
\end{align}
In this limit, the density matrix Eq \eqref{density-dependece} will be simplified to

\begin{align}
    \rho^{\mathrm{bW}}(V_R)= \frac{1}{N}
    \begin{pmatrix}
         0 & 0 & 0 & 0 & 0 & 0 \\
        0 & \frac{m_t^2 y^2}{2m_W^2}V_R^2  & 0 & \frac{m_t y^2}{\sqrt{2}m_W}V_R & 0 & 0 \\
        0 & 0 & y^2 V_R^2 & 0 & \frac{m_t y^2}{\sqrt{2}m_W}V_R & 0 \\
        0 & \frac{m_t y^2}{\sqrt{2}m_W}V_R & 0 & y^2 & 0 & 0 \\
        0 & 0 & \frac{m_ty^2}{\sqrt{2}m_W}V_R & 0 & \frac{m_t^2 y^2}{2m_W^2} & 0 \\
        0 & 0  & 0 & 0 & 0 & 0 \\
    \end{pmatrix},
\end{align}
The diagonal elements contain either the SM contribution or terms quadratic in $V_R$, while all terms linear in $V_R$ appear in the off-diagonal elements. Consequently, in the $m_b\to0$ limit, the top-quark decay rate and the $W$ boson helicity fractions have no linear sensitivity to $V_R$ and depend primarily on $V_R^2$. For finite $m_b$, linear interference contributions are present but are suppressed by the bottom-quark mass.

By contrast, the off-diagonal elements retain terms linear in $V_R$, which encode the coherence between different joint $bW$ helicity configurations. These terms vanish if either the $b$-quark spin or the $W$-boson polarization is traced over. Their linear sensitivity to $V_R$ is therefore accessible only through genuine joint $bW$ spin-correlation observables, rather than through the $b$-quark or $W$-boson polarization alone.

A similar pattern arises for the tensor couplings $g_L$ and $g_R$. 
In the conventional parametrization of the $Wtb$ vertex, the amplitude proportional to $g_R$ can interfere directly with the dominant SM amplitude without suppression by the $b$-quark mass, thereby generating potentially sizable linear corrections to both the decay rate and the $W$ polarization observables. 
In contrast, the interference between the amplitude proportional to $g_L$ and the SM contribution is suppressed in conventional observables in the $m_b\rightarrow0$ limit. Nevertheless, unsuppressed linear sensitivity to $g_L$ can survive in off-diagonal spin coherence terms when the complete $bW$ system is analyzed.

The different components of the spin density matrix therefore provide complementary probes of the anomalous couplings.
The diagonal elements determine the helicity populations and contribute directly to the polarization observables. 
The off-diagonal elements instead encode coherence between different helicity amplitudes and give rise to polarization and spin-correlation observables. 
As a result, anomalous interactions that induce similar changes in the decay rate or the $W$-boson helicity fractions can still be distinguished through their characteristic patterns across the full set of polarization and spin-correlation parameters.

\section{From Density Matrix to Observables.}\label{sec:matrix_to_observables}

In the previous section, we introduced the spin-density-matrix formalism as a model-independent framework to characterize the spin structure of the $bW$ system produced in the top-quark decay $t\rightarrow bW$.
The density matrix $\rho^{bW}$ contains the complete polarization information of the $b$ quark and the $W$ boson, as well as their spin correlations.
Unlike conventional measurements based mainly on the $W$-boson helicity fractions, the density matrix description provides access to additional spin-dependent observables involving the $b$ quark.
However, the density matrix itself is not directly observable.
Therefore, it is necessary to establish a connection between its components and the angular distributions of experimentally accessible final-state particles.

Experimentally, the spin information of the $b$ quark can be accessed through the hadronization process into heavy baryons, in particular the $\Lambda_b$ baryon, which retains a fraction of the original $b$-quark polarization~\cite{Kats:2023zxb,Afik:2025grr}.
In this work, we consider top-quark pair production at the HL-LHC and focus on the decay of one of the two top quarks via the decay chain
\begin{equation}
\begin{aligned}
pp \to t\bar t,\quad
t\to bW^+ ,\quad
W^+\to \ell^+ \nu_\ell,\quad
b\to \Lambda_b + X_{\rm had},\quad
\Lambda_b \to X_c \ell^- \bar\nu_\ell ,\\
pp \to t\bar t,\quad
\bar t\to \bar b W^- ,\quad
W^-\to \ell^- \bar\nu_\ell,\quad
\bar b\to \bar \Lambda_b + X_{\rm had},\quad
\bar \Lambda_b \to  X_c \ell^+ \nu_\ell,
\end{aligned}
\label{signalevent}
\end{equation}
where $X_c$ denotes a charmed baryon, such as $\Lambda_c^\pm$.
The angular distributions of the leptons from $\Lambda_b$ decay provide information on the $b$-quark spin, while the charged lepton from the leptonic decay of the $W$ boson acts as a spin analyzer of the $W$ boson.
Therefore, the combined angular information from the two decay chains allows the reconstruction of the complete $bW$ spin structure.

From a theoretical perspective, measuring the momenta of the final-state decay products can be interpreted as a generalized measurement of the spin state. For the $b$ quark, the spin projection operator along a unit vector $\hat{n}$ in the rest frame of the baryon is
\begin{align}
\Pi_b(\hat{n})=\frac{1}{2} I_2+\frac{1}{2} n_i \sigma_i.
\end{align}
Similarly, for the spin-1 $W$ boson, the projection operator is given by\cite{Bi:2023uop}
\begin{align}
\Pi_W(\hat{n})=\frac{1}{3} I_3+\frac{1}{2} n_i S_i+\frac{1}{4}\left(n_i n_j-\frac{1}{3} \delta_{i j}\right) S_{\{i j\}}.
\end{align}
This formalism allows one to systematically map the full density matrix onto the angular distributions of the measurable final-state particles.

Using these projection operators, the normalized differential cross section for observing the spin analyzers for the $b$ quark and the $W$ boson along specific directions can be written as
\begin{align}
\frac{\mathrm d \sigma}{\sigma \mathrm d \Omega_b \mathrm d \Omega_W}=\frac{1}{N_{b W}} \operatorname{Tr}\left[\rho_{b W} \Pi_b\left(\vec{n}_b\right) \otimes \Pi_W\left(\vec{n}_W\right)\right],
\end{align}
where $N_{b W}$ is a normalization factor. 
After substituting the explicit forms of the density matrix and the decay operators, the simplified result reveals a transparent decomposition
\begin{align}\label{eq:distri}
\frac{\mathrm d \sigma}{\sigma \mathrm d \Omega_b \mathrm d \Omega_W}=\frac{1}{(4 \pi)^2}\left(1+\tilde{B}_i n_i^b+3 \tilde{d}_j n_j^W+3 \tilde{q}_{i j} n_i^W n_j^W+3 \tilde{C}_{i j} n_i^b n_j^W+3 \tilde{C}_{j k}^i n_i^b n_j^W n_k^W\right),
\end{align}
where $n_i^{b/W}=\cos\theta_i^{b/W}$ denote the direction cosines of the corresponding spin analyzers in the respective parent rest frames with respect to the helicity axes $\{\hat n,\hat r,\hat k\}$.
In this work, we adopt the helicity basis with the longitudinal direction $\hat k$ defined by the momentum direction of the $b$ quark.
The coefficients in Eq.~(\ref{eq:distri}) correspond directly to measurable spin observables.

In realistic measurements, the experimentally extracted observables are affected by the spin-analyzing power of the decay products, the polarization retention of the $\Lambda_b$ baryon, and the purity of the selected event sample.
The effective observable coefficients are related to the density matrix parameters by
\begin{equation}
B_i=\alpha_b r_i f \tilde{B}_i,\quad C_{ij}=\alpha_b\alpha_Wr_i f\tilde{C}_{ij}.
\end{equation}
where $\alpha_b$ and $\alpha_W$ denote the spin-analyzing powers of the chosen decay products in the $\Lambda_b$ decay and the $W$-boson decay, respectively.
The charged lepton from the leptonic decay of the $W^+$ boson provides an optimal spin analyzer with $\alpha_{W^+}\simeq1$, while the neutrino in the $\Lambda_b$ decay chain allows access to the original $b$-quark polarization with $\alpha_b\simeq1$~\cite{Manohar:1993qn,Galanti:2015pqa,Kats:2023zxb}.
For the corresponding antiparticle decays, the spin-analyzing powers acquire opposite signs, $\alpha_{\bar b}\simeq -1$, $\alpha_{W^-}=-1$. The polarization retention factors $r_i$ quantify the fraction of the original $b$-quark polarization retained by the $\Lambda_b$ baryon after fragmentation.
They can be regarded as effective parameters describing spin-dependent fragmentation, with values depending on the polarization direction.
Typical ranges are
\begin{align}
0.4 \lesssim r_L \lesssim 0.8, \quad 0.5 \lesssim r_T \lesssim 0.8,
\end{align}
for the longitudinal $(\bm{\hat{k}})$ and transverse $(\bm{\hat{n}},\bm{\hat{r}})$ components, respectively.
The sample purity is defined as
\begin{align}
f=\frac{N_{\rm sig}}{N_{\rm sam}},
\end{align}
where $N_{\rm sig}$ and $N_{\rm sam}$ denote the number of signal events and the total number of selected events, respectively.

The density-matrix elements can be obtained by evaluating appropriate expectation values over the angular distributions~\cite{Bi:2023uop}, leading to the relations: 
\begin{equation}
\begin{array}{ll}
B_i=3\left\langle n_i^b\right\rangle, &d_i=\left\langle n_i^W\right\rangle, \qquad q_{ij}=\frac{5}{2}\left\langle n_{i}^Wn_j^W-\frac{1}{3}\delta_{ij}\right\rangle, \\
C_{ij}=3\left\langle n_i^bn_j^W\right\rangle, &
C^i_{jk}=\frac{15}{2}\left\langle n_i^b\left(n_j^Wn_k^W-\frac{1}{3}\delta_{jk}\right)\right\rangle.
\end{array}
\end{equation}

These relations provide a direct and model-independent mapping between the theoretical description of the $bW$ density matrix and experimentally measurable angular correlations.
By analyzing the angular distributions of the neutrino from $\Lambda_b$ decay and the charged lepton from the $W$-boson decay, we can extract the polarization of the $b$ quark, the polarization of the $W$ boson, and the spin correlations between them.
These spin observables provide complementary information to conventional top-quark measurements and enable sensitive probes of anomalous $Wtb$ interactions at the HL-LHC.

\section{Statistical Sensitivity of Spin Observables} \label{sec:uncertainty}

In the previous section, we established the connection between the $bW$ density matrix and experimentally measurable angular observables.
To evaluate the sensitivity of these observables to anomalous $Wtb$ interactions, we estimate their statistical uncertainties based on the expected angular distributions of the $\Lambda_b$ and $W$-boson decay products.

\subsection{Observable distributions}

The spin observables introduced in Sec.~\ref{sec:matrix_to_observables} are extracted from the normalized differential angular distribution,
\begin{equation}
\frac{\mathrm dp}
{\mathrm d\Omega_b\mathrm d\Omega_W}
\equiv
\frac{1}{\sigma}
\frac{\mathrm d\sigma}
{\mathrm d\Omega_b\mathrm d\Omega_W},
\end{equation}
where $\mathrm d\Omega_i=\mathrm d\cos\theta_i\,\mathrm d\varphi_i$ $(i=b,W)$. Since the full angular distribution contains contributions from multiple spin observables simultaneously, the extraction of an individual observable requires an appropriate projection of the multidimensional distribution.

For a given observable $c$, Eq.~(\ref{eq:distri}) can be expressed schematically as
\begin{equation}
\frac{\mathrm dp}{\mathrm d\Omega_b \mathrm d\Omega_W}=\frac{1}{(4\pi)^2}\left[1+c\,U(\hat n_b,\hat n_W)+\sum_\alpha c_\alpha U_\alpha (\hat n_b,\hat n_W)\right],
\end{equation}
where $U(\hat n_b,\hat n_W)$ represents the angular structure associated with the observable of interest, while the remaining terms describe the contributions of other spin observables to the same distribution.

To isolate the sensitivity to $c$, we introduce a new variable $u= U(\hat n_b,\hat n_W)$, supplemented by a set of auxiliary variables $\vec v= \vec V(\hat n_b,\hat n_W)$. The probability distribution then becomes
\begin{equation}
\frac{\mathrm dp}{\mathrm du\mathrm d\vec v}=\frac{1}{(4\pi)^2}\left[1+c u+\sum_\alpha c_\alpha U'_\alpha(u,\vec v)\right]J(u,v),
\end{equation}
where the Jacobian determinant for variable transformation is
\begin{equation}
J(u,\vec v)\equiv\left|\frac{\partial(\hat n_b,\hat n_W)}{\partial(u,\vec v)}\right|.
\end{equation}
After integrating over the auxiliary variables, the distribution of the single observable $u$ is obtained as
\begin{equation}
\frac{\mathrm dp}{\mathrm du}=\frac{1}{(4\pi)^2}\left[ \left(1+cu\right)\int J(u,\vec v)\,\mathrm d\vec v+\sum_\alpha c_\alpha\int U'_\alpha(u,\vec v)J(u,\vec v)\,\mathrm d\vec v \right].
\end{equation}
This procedure provides a direct way to quantify the statistical sensitivity associated with each spin observable.

\subsection{Expected uncertainties}

For a conservative estimate of the statistical sensitivity, we vary only the parameter of interest while holding all other observable coefficients fixed, namely $\Delta c_\alpha=0$. Following Ref.~\cite{Kats:2023zxb}, we parameterize the projected
one-dimensional distribution as
\begin{equation}
\frac{\mathrm dp}{\mathrm du}
=
\frac{1}{\mathrm{Norm}}
(1+cu)\,f(u).
\end{equation}
The statistical uncertainty on the observable $c$ can then be expressed as
\begin{equation}
\Delta c
=
\frac{1}{S(c)\sqrt{N_{\rm eff}}},
\end{equation}
where $N_{\rm eff}$ denotes the number of effective signal events and
\begin{equation}
S^2(c)
\simeq
\frac{1}{\mathrm{Norm}}
\int
\frac{u^2}{1+cu}
f(u)\mathrm du
\end{equation}
is the sensitivity function. After using the following relationships:
\begin{equation}
f(u)=\int J(u,\vec v)\,\mathrm d\vec v,
\qquad
\mathrm{Norm}=(4\pi)^2,
\end{equation}
the sensitivity function can be directly evaluated from the angular structure of the corresponding observable,
\begin{equation}
S^2(c)
\simeq
\frac{1}{(4\pi)^2}
\int
\frac{
U(\hat n_b,\hat n_W)^2
}{
1+c\,U(\hat n_b,\hat n_W)
}
\,\mathrm d\Omega_b \mathrm d\Omega_W.
\end{equation}

In the following analysis, we evaluate the sensitivity function at $c=0$, which provides a conservative estimate of the expected statistical precision, $S(c)|_{c=0}$.
The resulting sensitivities for different spin observables are summarized in Tab.~\ref{tab:uncertainty} 
, which presents the angular structures associated with each observable, the corresponding sensitivity functions, and the resulting statistical uncertainties.

\begin{table}[h] 
\centering 
\setlength{\tabcolsep}{20pt}
\renewcommand{\arraystretch}{1.5}
\begin{tabular}{|c|c|c|c|}\hline 
$c$ & $U(\hat n_b,\hat n_W)$ & $S^2(c)|_{c=0}$ & $\Delta c$\\\hline 
$\alpha_b r_i B_i$ & $\cos\theta_i^b$ & $\frac{1}{3}$ & $\Delta B_i=\frac{\sqrt{3}}{r_i\alpha_b}\frac{1}{\sqrt{N}}$\\\hline 
$3\alpha_W d_i$ & $\cos\theta_i^W$ & $\frac{1}{3}$ & $\Delta d_i=\frac{1}{3}\frac{\sqrt{3}}{\alpha_W}\frac{1}{\sqrt{N}}$\\\hline 
$3\alpha_b\alpha_W r_iC_{ij}$ & $\cos\theta_i^b\cos\theta_j^W$ & $\frac{1}{9}$ & $\Delta C_{ij}=\frac{1}{3}\frac{3}{r_i\alpha_b\alpha_W}\frac{1}{\sqrt{N}}$\\\hline 
$3\alpha_W^2 q_{ii}$ & $\frac{3}{2}\cos^2\theta_i^W-\frac{1}{2}$ & $\frac{1}{5}$ & $\Delta q_{ii}=\frac{1}{3}\frac{\sqrt{5}}{\alpha_W^2}\frac{1}{\sqrt{N}}$\\\hline
$3\alpha_W^2q_{ij}(i\neq j)$ & $ \cos\theta_i^W\cos\theta_j^W $ & $\frac{1}{15}$ & $\Delta q_{ij}=\frac{1}{3}\frac{\sqrt{15}}{\alpha_W^2}\frac{1}{\sqrt{N}}$\\\hline 
$3\alpha_b\alpha_W^2 C^i_{jj}$ & $\cos\theta_i^b(\frac{3}{2}\cos^2\theta_j^W-\frac{1}{2})$ & $\frac{1}{15}$ & $\Delta C^i_{jj}=\frac{1}{3}\frac{\sqrt{15}}{r_i\alpha_b\alpha_W^2}\frac{1}{\sqrt{N}}$\\\hline 
$3\alpha_b\alpha_W^2 C^i_{jk}(j\neq k)$ & $ \cos\theta_i^b\cos\theta_j^W\cos\theta_k^W $ & $\frac{1}{45}$ & $\Delta C^i_{jk}=\frac{1}{3}\frac{\sqrt{45}}{r_i\alpha_b\alpha_W^2}\frac{1}{\sqrt{N}}$\\\hline 
\end{tabular} 
\caption{The single variable $U(\hat n_b,\hat n_W)$, integral results for $S^2(0)$ and the estimation of uncertainties $\Delta c$ for different observables.} \label{tab:uncertainty} 
\end{table}

For observables involving tensor structures, such as $q_{ij}$ and $C^i_{jk}$, the symmetry and traceless conditions of the corresponding density-matrix components must be taken into account. For example, when considering $q_{zz}$, we have
\begin{equation}
q_{xx}+q_{yy}=-q_{zz}.
\end{equation}
After averaging over $\varphi_W$, the corresponding angular dependence can
be expressed as
\begin{equation}
q_{zz}
\left(
\frac32\cos^2\theta_W-\frac12
\right),
\end{equation}
and therefore the same expression can be used for the diagonal tensor observables. Similar considerations apply to $C^i_{jj}$.

\section{Event Analysis}
\label{sec:analysis}

We investigate the sensitivity to anomalous $Wtb$ couplings at the HL-LHC with a center-of-mass energy $\sqrt{s}=14~\rm TeV$ and an integrated luminosity $\mathcal{L}_{\rm HL-LHC} = 3000 ~\rm fb^{-1}$.
The large production rate of top-quark pairs at the HL-LHC provides an ideal environment for precision studies of top-quark spin information.
In particular, the combination of the leptonic decay of the $W$ boson and the semileptonic decay of the $\Lambda_b$ baryon allows us to access both the $W$-boson polarization and the $b$-quark spin information introduced in Sec.~\ref{sec:matrix_to_observables}.

\subsection{Signal process and event reconstruction}

We consider $t\bar t$ production in which the $bW$ spin system is reconstructed from the decay products of one top quark, while the other top quark is required to decay hadronically. For the $bW$ spin system from leptonic decay of top, the $W$ boson decays leptonically, with the charged lepton serving as a spin analyzer of the $W$ boson. The accompanying $b$ quark hadronizes into a $\Lambda_b$ baryon, whose semileptonic decay provides access to the spin information inherited from the parent $b$ quark. The signal processes are therefore
\begin{align}
pp \to t\bar t,\qquad
\left(t\to bW^+ ,
b\to \Lambda_b + X_{\rm had},
\Lambda_b \to X_c \mu^- \bar\nu_\mu , W^+\to \ell^+ \nu_\ell \right) \left( \bar t \to \bar b j j\right),\\
pp \to t\bar t,\qquad
\left(\bar t\to \bar b W^- , 
\bar b\to \bar \Lambda_b + X_{\rm had},
\bar \Lambda_b \to  X_c \mu^+ \nu_\mu , W^-\to \ell^- \bar\nu_\ell \right) \left( t\to b jj\right),
\end{align}
This topology provides simultaneous access to the spin information of both constituents of the $bW$ system. The charged lepton from the $W$-boson decay probes the $W$ polarization, while the angular distribution of the neutrino from  $\Lambda_b$ decay carries information about the $b$-quark polarization. Their joint angular distribution therefore provides the experimental input for extracting the $bW$ polarization and spin-correlation observables introduced in Sec.~\ref{sec:matrix_to_observables}. Requiring the opposite-side top quark to decay hadronically avoids an additional neutrino from the corresponding $W$ decay and provides additional kinematic constraints for event reconstruction. The isolated charged lepton from the signal-side $W$ decay also enables efficient triggering and helps suppress QCD multijet backgrounds.

For each selected event, the $bW$ spin system top-quark candidate is reconstructed from the identified $b$ jet and the corresponding $W$-boson decay products. 
The momentum of the neutrino from the leptonic $W$ decay is constrained using the missing transverse momentum together with the on-shell condition
\begin{equation}
m_W^2=(p_\ell+p_\nu)^2,
\end{equation}
and the reconstructed $W$-boson momentum is combined with the $b$-jet momentum to obtain
\begin{equation}
p_t=p_b+p_W.
\end{equation}
The reconstructed event kinematics are then transformed to the reference frames required to evaluate the angular variables $n_i^b$ and $n_i^W$. The helicity basis, angular observables, and $\Lambda_b$ polarization-retention effects have been defined in Sec.~\ref{sec:matrix_to_observables} and are not repeated here.

Because the signal-side decay chain contains two neutrinos, one from the $W$ decay and one from the semileptonic $\Lambda_b$ decay, their combined contribution to the missing transverse momentum introduces an additional ambiguity in the event reconstruction. In the present sensitivity study, the associated reconstruction effects are absorbed into the overall signal efficiency. The resulting event-selection efficiency, background contamination, and effective signal yield are discussed below.

\subsection{Signal yield and background estimation}
The number of selected top-quark pair events in the semileptonic decay channel can be calculated using
\begin{equation}
N_{\rm sig }^{t\bar t}
=
\mathcal{L}_{\rm HL-LHC}
\epsilon_{\rm eff}^{t\bar t} \sigma_{t\bar t} ,
\end{equation}
where $\sigma_{t\bar t}=985.7~\rm pb$ ~\cite{Czakon:2013goa,ParticleDataGroup:2024cfk} is the top-quark pair production cross section at $\sqrt{s}=14~\rm TeV$. 
The selection efficiency $\epsilon_{\rm eff}^{t\bar t}$ is estimated based on existing LHC experimental analyses~\cite{ATLAS:2020aln,CMS:2021vhb} and extrapolated to the HL-LHC scenario. 
The resulting efficiency accounts for the semileptonic topology, in which one $W$ boson decays leptonically and the other hadronically, as well as the associated reconstruction and event selection criteria.

To measure the $b$-quark spin information, we select signal events in which a $b$ quark hadronizes into a $\Lambda_b$ baryon. For the inclusive selection, the expected number of signal events is
\begin{equation}
N_{\rm sig}^{\rm incl}
=
N_{\rm sig }^{t\bar t}
\cdot
f(b\rightarrow\Lambda_b)
\cdot
{\rm BR}(\Lambda_b\rightarrow X_c\mu^-\bar\nu_\mu).
\end{equation}
A more restrictive exclusive selection is obtained by requiring the explicit reconstruction of the charmed baryon,
\begin{equation}
N_{\rm sig}^{\rm excl}
=
N_{\rm sig }^{t\bar t}
\cdot
f(b\rightarrow\Lambda_b)
\cdot
{\rm BR}(\Lambda_b\rightarrow X_c\mu^-\bar\nu_\mu)
\cdot
{\rm BR}(\Lambda_c^+\rightarrow{\rm reco.})
\cdot
\epsilon_{\rm reco}.
\end{equation}
We adopt the numerical inputs for the fragmentation fraction, branching fractions, and reconstruction efficiency given in Ref.~\cite{Kats:2023zxb},
\begin{align}
&f(b\rightarrow\Lambda_b)\simeq 7\% , \nonumber\\
&{\rm BR}(\Lambda_b\rightarrow X_c\mu^-\bar\nu_\mu)
\simeq 11\% , \nonumber\\
&{\rm BR}(\Lambda_c^+\rightarrow{\rm reco.})
\simeq18\% , \nonumber\\
&\epsilon_{\rm reco}=50\%.
\end{align}

We consider the backgrounds from two different classes~\cite{ATLAS:2020aln,Kats:2023zxb}: extrinsic and intrinsic backgrounds. Extrinsic backgrounds arise from processes other than the signal $t\bar t$ production that nevertheless satisfy the event-level selection and mimic the reconstructed semileptonic $t\bar t$ topology. Intrinsic backgrounds occur within events containing a genuine $b$ jet and arise when other heavy-flavor decays mimic $\Lambda_b$ decay used to analyze the $b$-quark spin.

\begin{table}[htbp]
    \centering
    \small
    \setlength{\tabcolsep}{12pt}
    \renewcommand{\arraystretch}{1.3}
    \begin{tabular}{|c|c|c|c|c|c|c|c|}\hline
        \multirow{2}{*}{Category} & Signal &\multicolumn{4}{|c|}{Backgrounds} \\\cline{2-6}
         & $N_{\rm sig}^{t\bar{t}}$ & Single top & Multijet & $W/Z$+jets \& diboson & $t\bar{t}X$   \\\hline
        \multirow{2}{*}{Event Number} & \multirow{2}{*}{$2.5\times10^7$} & $9.5\times 10^5$& $5.6\times10^5$ & $5.8\times 10^{5}$ & $1.8\times 10^{5}$  \\\cline{3-6}
        &  &\multicolumn{4}{|c|}{$N_\mathrm{bkg}^{\rm extr}=2.3\times10^6$} \\\hline
    \end{tabular}
    \caption{Expected event yields for the $t\bar{t}$ signal and backgrounds.}
    \label{tab:ttbk}
\end{table}

Extrinsic backgrounds originate from processes other than signal $t\bar t$ production and contribute to the selected sample by mimicking the reconstructed semileptonic $t\bar t$ topology.
The event selection requires at least five reconstructed jets, exactly two of which are $b$ tagged. After this selection, the dominant backgrounds arise from single-top production, multijet events, $W / Z+$ jets and diboson contribution, and $t \bar{t} X$ production. The expected yields of the selected $t\bar t$ signal and these four extrinsic background categories are summarized in Tab.~\ref{tab:ttbk}.
Denoting the total contribution by $N_{\rm bkg}^{\text {extr }}$, we define the corresponding event-level purity as
\begin{equation}
f_{\mathrm{extr}}^{t \bar{t}}=\frac{N_{\rm sig}^{t \bar{t}}}{N_{\rm sig}^{t \bar{t}}+N_{\rm bkg}^{\mathrm{extr}}} .
\end{equation}

The intrinsic backgrounds enter at the subsequent $\Lambda_b$-reconstruction stage. The desired signal contains a genuine decay chain
\begin{equation}
b \rightarrow \Lambda_b+X, \quad \Lambda_b \rightarrow X_c \mu^{-} \bar{\nu}_\mu,
\end{equation}
whereas semileptonic decays of other $b$ hadrons can also produce a displaced muon inside the $b$ jet and therefore mimic the $\Lambda_b$ signal. We characterize this contamination by the intrinsic purity
\begin{equation}
f_{\mathrm{intr}}^{\rm incl (excl)}=\frac{N^{\rm incl (excl)}_{\mathrm{sig}}}{N^{\rm incl (excl)}_{\mathrm{sig}}+N_{\rm bkg}^{\mathrm{intr}}},
\end{equation}
where $N_{\rm bkg}^{\text {intr }}$ denotes the number of background events from other heavy-flavor decays that satisfy the same $\Lambda_b$ reconstruction requirements.

For the inclusive reconstruction, only a displaced muon inside the $b$ jet is required. This maximizes the signal efficiency, but also retains a large contribution from ordinary semileptonic $B$-meson decays.
The intrinsic purity is estimated from the relative fragmentation fractions and semileptonic branching fractions of the $\Lambda_b$ signal and the competing $b$-hadron decays~\cite{Kats:2023zxb}. This gives
\begin{equation}
f_{\text {intr }}^{\text {incl }} \simeq 7.4 \% .
\end{equation}
For the exclusive reconstruction, a $\Lambda_c^{+}$ decay is explicitly reconstructed in addition to the displaced muon. This requirement strongly suppresses ordinary semileptonic $B$-meson backgrounds, although residual contamination from $B$-meson decays containing a genuine $\Lambda_c^{+}$remains.
We adopt the following estimate for the purity~\cite{Kats:2023zxb}:
\begin{equation}
f_{\text {intr }}^{\text {excl }} \simeq 66 \% .
\end{equation}

We next incorporate these background contributions into the statistical sensitivity analysis of the spin observables. Consider a generic sample containing $N_{\text {sig }}$ signal events and $N_{\text {bkg }}$ background events. Its signal purity is
\begin{align}
f=\frac{N_{\mathrm{sig}}}{N_{\mathrm{sig}}+N_{\mathrm{bkg}}} .
\end{align}

In our analysis, the background dilution occurs at two successive stages. The extrinsic purity describes the contamination of the event-level $t \bar{t}$ sample, whereas the intrinsic purity describes the contamination of the subsequently reconstructed $\Lambda_b$ sample. Assuming that these two effects approximately factorize, the total purity can be written as
\begin{equation}
f^{\rm incl (excl)}=f_{\text {extr }}^{t \bar t} f_{\mathrm{intr}}^{\rm incl (excl)}.  
\end{equation}
The effective event yield entering the statistical sensitivity is consequently given by
\begin{equation}
N_{\text {eff }}^{\rm incl (excl)}=N_{\text {sig }}^{\rm incl (excl)} \times f^{\rm incl (excl)}.
\end{equation}
Under the factorization assumption stated above, the total purity is obtained by combining the event-level purity $f_{\text {extr }}^{t \bar{t}}=92 \%$ with the intrinsic purity of each reconstruction strategy. 
For the exclusive reconstruction, we obtain
\begin{equation}
f^{\mathrm{excl}}=0.92 \times 0.66 \simeq 0.60,
\end{equation}
and hence
\begin{align}
N_{\mathrm{eff}}^{\mathrm{excl}}=f^{\mathrm{excl}} N_{\mathrm{sig}}^{\mathrm{excl}} \simeq 0.60 \times 8700=5220 .
\end{align}
For the inclusive reconstruction, the intrinsic purity is lower, resulting in a total purity of
\begin{align}
f^{\text {incl }}=0.92 \times 0.074 \simeq 6.8 \%,
\end{align}
which gives
\begin{align}
N_{\mathrm{eff}}^{\mathrm{incl}}=f^{\mathrm{incl}} N_{\mathrm{sig}}^{\mathrm{incl}} \simeq 0.068 \times 9.7\times10^4=6596 .
\end{align}
The resulting signal yields, purities, and effective event numbers are summarized in Tab. ~\ref{tab:ttevts_num}.

Although the exclusive reconstruction achieves a substantially higher intrinsic purity, $f_{\text {intr }}^{\text {excl }}=66 \%$, its stronger reconstruction requirements reduce the signal yield. Conversely, the inclusive reconstruction has a much lower intrinsic purity, $f_{\text {intr }}^{\text {incl }}=7.4 \%$, but retains a significantly larger signal sample. 
At the purely statistical level considered here, the larger inclusive signal yield compensates for the stronger background dilution, resulting in $N_{\mathrm{eff}}^{\mathrm{incl}}=6596$, compared with $N_{\mathrm{eff}}^{\mathrm{excl}}=5220$.
Consequently, the statistical uncertainty for the exclusive selection is larger than that for the inclusive selection by a factor of $\sqrt{6596/5220}\simeq 1.1$. We therefore adopt the inclusive reconstruction as the benchmark for the statistical sensitivity estimates presented below. For the charge-conjugate antitop decay, we assume the same effective event yield.

For the present projection, we use the effective yield, $N_{\mathrm{eff}}$, to characterize the statistical sensitivity of the signal-plus-background sample. This amounts to replacing the background-contaminated sample by an equivalent pure signal sample with the same statistical sensitivity, allowing the analysis to be expressed directly in terms of the $bW$ polarization and spin-correlation coefficients without explicitly modeling the background angular distributions. The background dilution is thus encoded in $N_{\mathrm{eff}}$ and the resulting statistical uncertainties.

This treatment assumes that the backgrounds do not generate the correlated $bW$ angular structures of interest, so that their dominant effect is to dilute the signal coefficients. Nontrivial background angular distributions and systematic uncertainties associated with background modeling and subtraction are not included in the present projection and would need to be treated explicitly in a full experimental analysis.

\begin{table}
    \centering
    \setlength{\tabcolsep}{8 pt}
    \renewcommand{\arraystretch}{1.3}
    \begin{tabular}{|c|c|c|c|c|c|c|c|c|}\hline
        Scheme & $N_{\rm sig}^\mathrm{t\bar t}=\mathcal{L}\sigma_\mathrm{t\bar t}\epsilon_\mathrm{cut}$ & $N_\mathrm{sig}^{\rm incl (excl)}$ & $N_\mathrm{bkg}^{\rm extr}$ & $f^\mathrm{t\bar t}_{\rm extr}$ & $f_{\mathrm{intr}}^{\rm incl (excl)} $& $f^{\rm incl (excl)}$ & $N_{\rm eff}^{\rm incl (excl)}$ \\\hline
        Exclusive & \multirow{2}{*}{$1.3\times10^7$} & 8700 & \multirow{2}{*}{$1.1\times10^6$} & \multirow{2}{*}{$92\%$} & 66\% & 60\% & 5220 \\ \cline{1-1} \cline{3-3}\cline{6-8}
        Inclusive & & $9.7\times 10^4$&  &   & 7.4\% & 6.8\% & 6596 \\\hline
    \end{tabular}
    \caption{Values used to estimate the effective number of signal events expected from top-quark decay in $t\bar{t}$ production at the HL-LHC.}
    \label{tab:ttevts_num}
\end{table}

\section{Results at the HL-LHC}
\label{sec:result}

Using the event yields and statistical uncertainties obtained in the previous sections, we now evaluate the sensitivity of the complete $b W$ spin-density matrix to anomalous $W t b$ couplings at the HL-LHC. We consider the anomalous couplings
\begin{align}
d=\left\{V_R, g_L, g_R\right\},
\end{align}
allowing both their real and imaginary parts to vary. For the one-dimensional analyses discussed below, one coupling component is varied at a time while all remaining anomalous couplings are fixed to their SM values.

For each coupling, the expected constraints are obtained from the $\chi^2$ statistic,
\begin{equation}
\chi^2(d)=\sum_i\left[
\left(
\frac{O_i^t(d)-O_i^{t,\rm SM}}
{\Delta O_i^t}
\right)^2 + \left(
\frac{O_i^{\bar t}(d)-O_i^{\bar t, \rm SM}}
{\Delta O_i^{\bar t}}
\right)^2 \right],
\end{equation}
where $O_i^t(d)$ and $O_i^{\bar{t}}(d)$ denote the spin observables predicted for top and antitop decays, respectively, and $\Delta O_i^{t, \bar{t}}$ are their corresponding statistical uncertainties. The sum runs over the measurable observables of the $b W$ spin-density matrix,
\begin{align}
\{B_i, d_i,  q_{ij},  C_{ij},  C^i_{jk}\}.
\end{align}
The top and antitop samples are treated as statistically independent contributions to the $\chi^2$.

For the numerical analysis, we adopt $|\alpha_b|=1,~ |\alpha_W|=1$,
and use a conservative polarization retention factor $r_i\simeq0.5 $.
For top-quark pair production, the effective event yields obtained from the HL-LHC event analysis are $N_t \simeq N_{\bar t} \simeq 6600$. With these inputs, the statistical uncertainties of the different classes of spin observables are given by
\begin{align}
\begin{gathered}
\Delta \tilde{B}_i=\frac{3.46}{\sqrt{N_i}}, \quad
\Delta \tilde{C}_{i i}=\frac{2}{\sqrt{N_i}}, \quad
\Delta \tilde{d}_i=\frac{0.58}{\sqrt{N_i}}, \\
\Delta \tilde{q}_{i i}=\frac{0.75}{\sqrt{N_i}}, \quad \Delta \tilde{q}_{i j}=\frac{1.29}{\sqrt{N_i}} \quad(i \neq j),\\
\Delta \tilde{C}_{j j}^i=\frac{2.58}{\sqrt{N_i}}, \quad \Delta \tilde{C}_{j k}^i=\frac{4.47}{\sqrt{N_i}} \quad(j \neq k),
\end{gathered}
\end{align}
with $N_i=N_{\mathrm{eff,t}}^{\mathrm{incl}}$ or $N_{\mathrm{eff,\bar{t}}}^{\mathrm{incl}}$ for the corresponding top or antitop sample.

\begin{figure}
    \centering
    \includegraphics[width=0.48\linewidth]{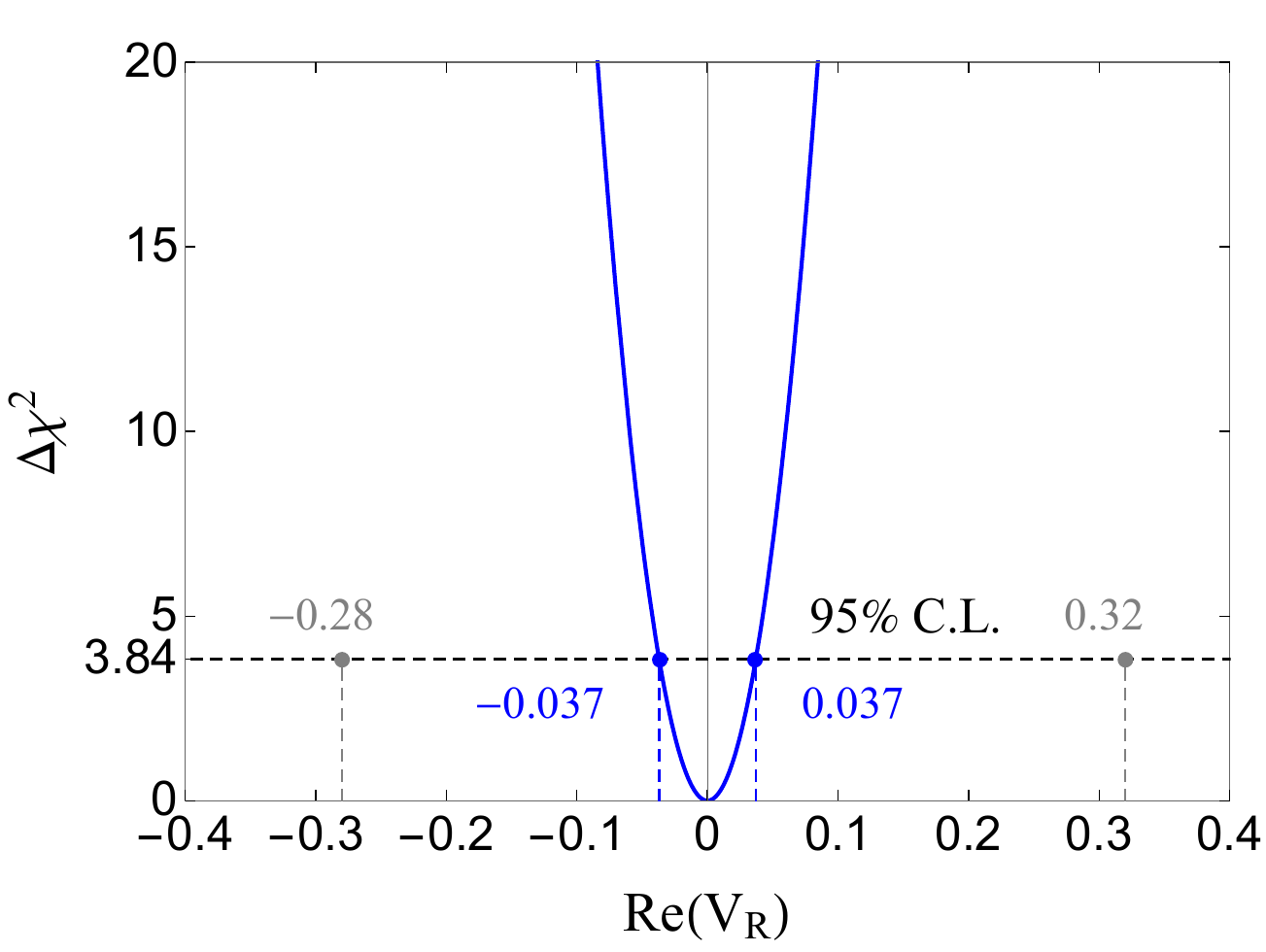}
    \includegraphics[width=0.48\linewidth]{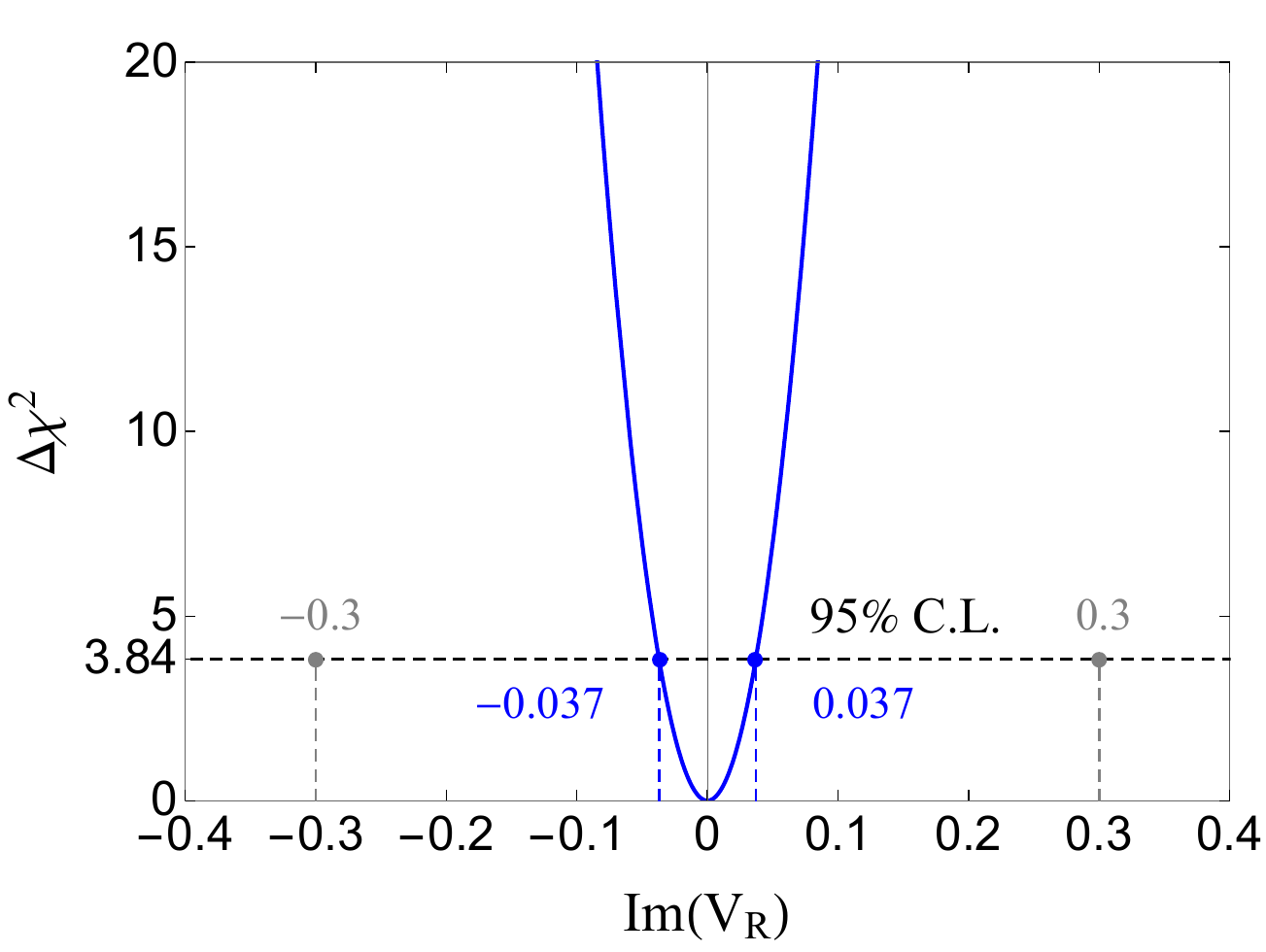}
    \includegraphics[width=0.48\linewidth]{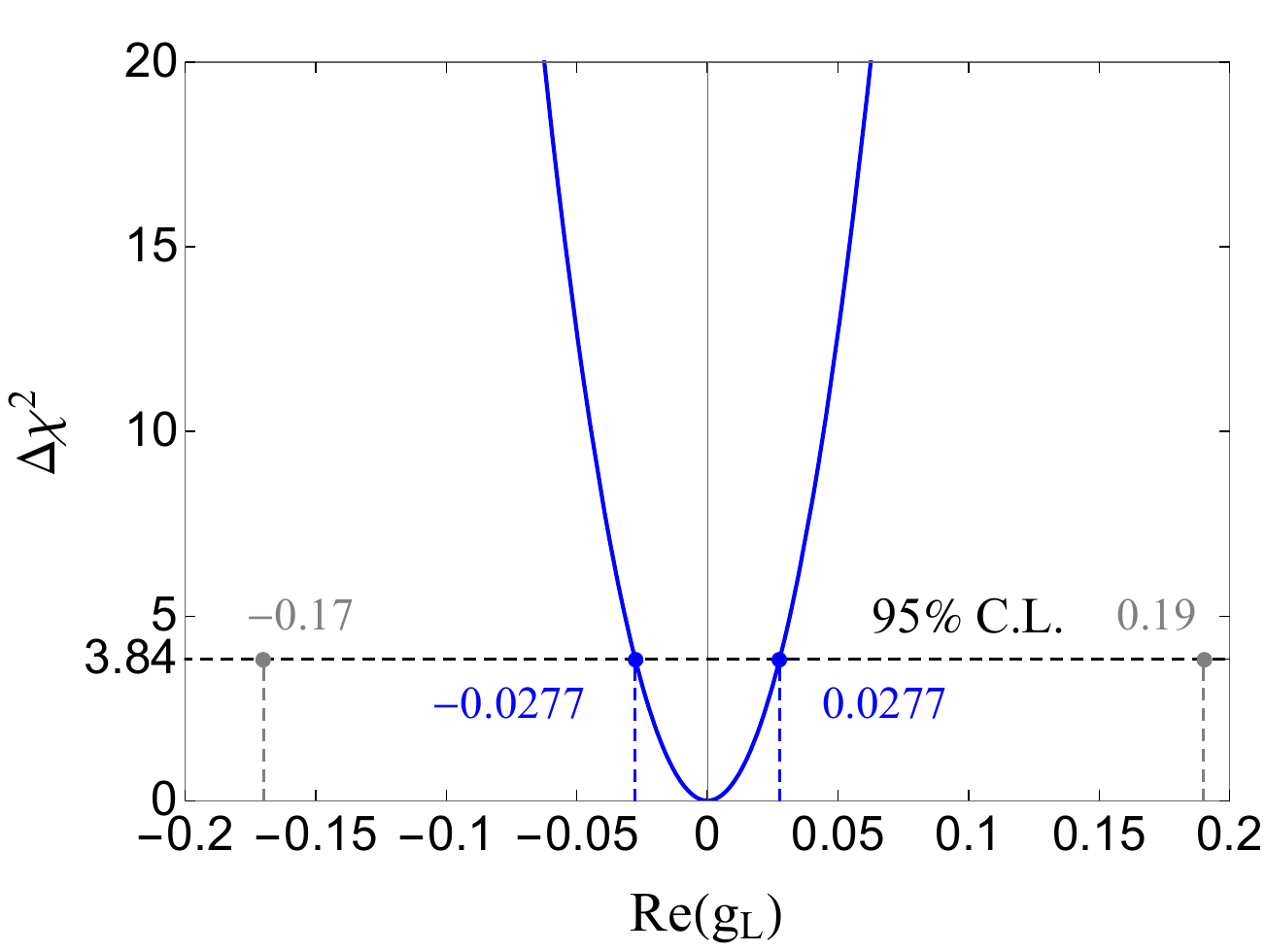}
    \includegraphics[width=0.48\linewidth]{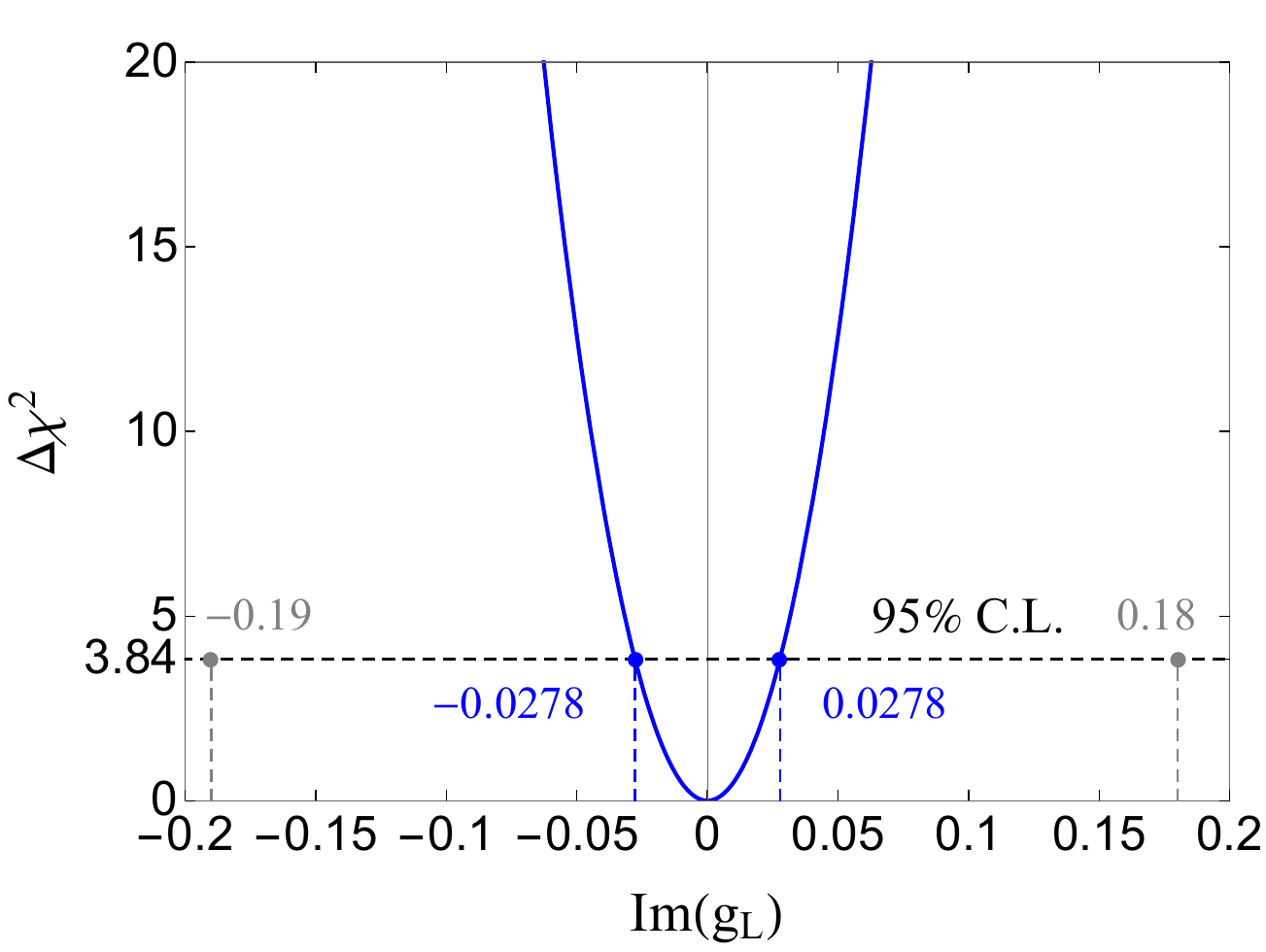}
    \includegraphics[width=0.48\linewidth]{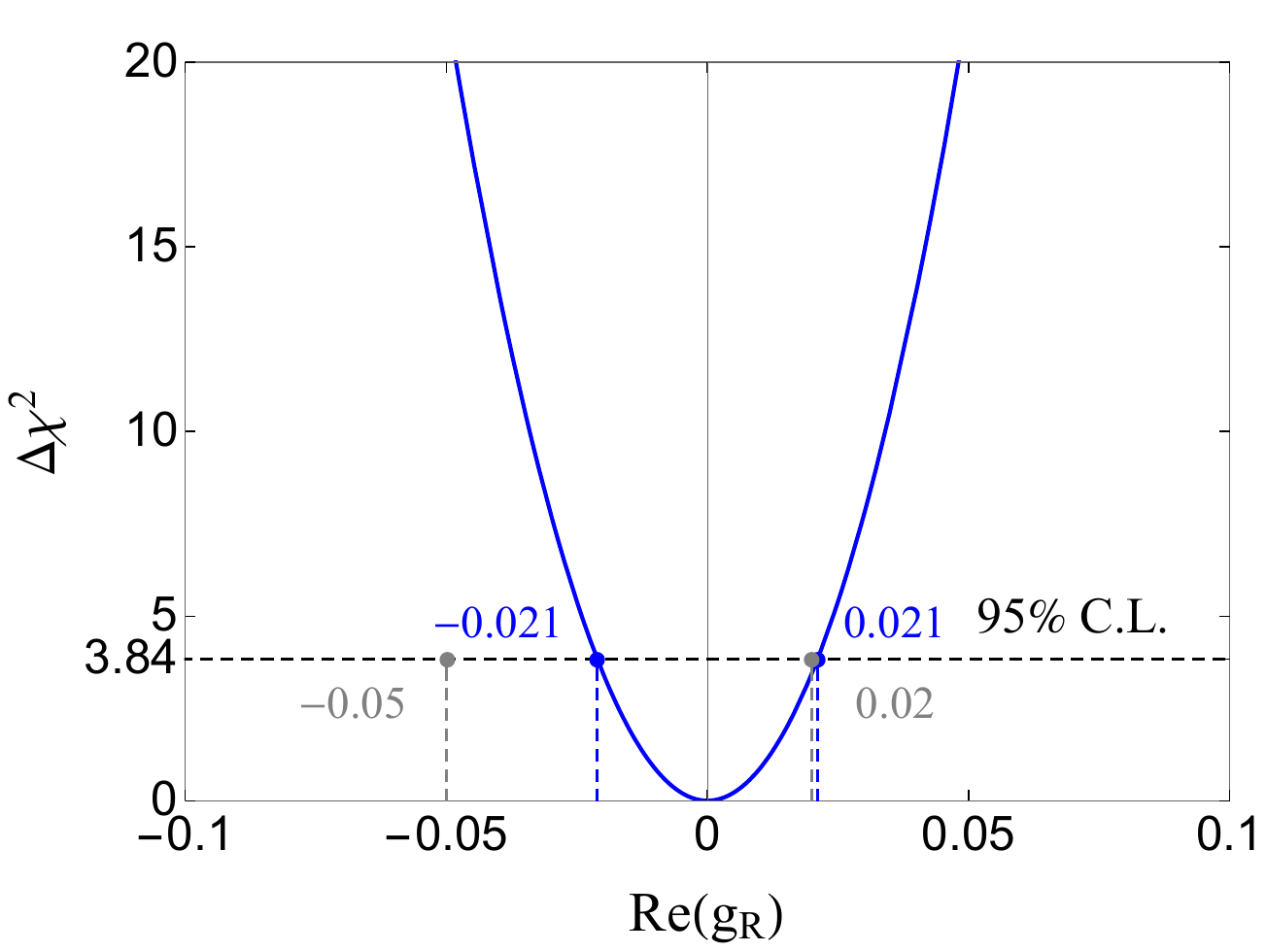}
    \includegraphics[width=0.48\linewidth]{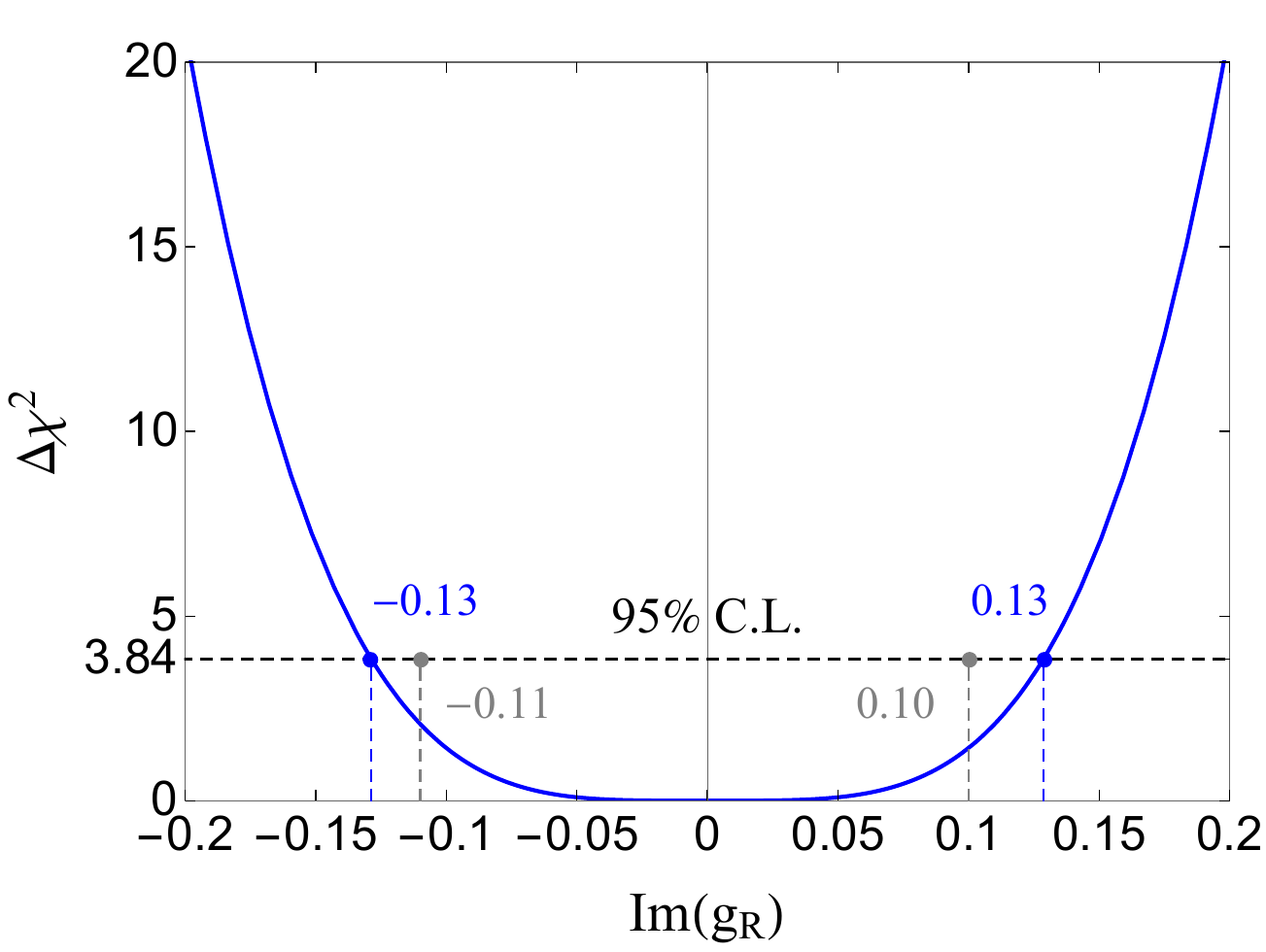}
    
\caption{The expected statistical significance $\chi^2$ as a function of new physics parameters $V_R,g_L,g_R$ in top-quark pair production.
To compare with the result in previous paper\cite{Deliot:2018jts}, we also plot their limits as the gray points. }
\label{fig:chi2_line}
\end{figure}
 
We first perform one-dimensional scans of the real and imaginary parts of $V_R, g_L$, and $g_R$. The expected constraints are obtained by varying the anomalous couplings around the SM point and evaluating the corresponding $\chi^2$ function. The resulting $\chi^2$ profiles are shown in Fig.~\ref{fig:chi2_line}. Since the SM prediction is taken as the reference point in the projected analysis, the minimum satisfies $\chi_{\min }^2=0 $. The expected 95\% confidence intervals for one parameter are therefore determined by
\begin{align}
\Delta \chi^2 \equiv \chi^2-\chi_{\min }^2=\chi^2=3.84 .
\end{align}
The resulting 95\% confidence intervals are
\begin{align}
\begin{array}{ll}
\operatorname{Re} (V_R) \in[-0.037,0.037], & \operatorname{Im} (V_R) \in[-0.037,0.037], \\
\operatorname{Re} (g_L) \in[-0.028,0.028], & \operatorname{Im} (g_L) \in[-0.028,0.028],\\
\operatorname{Re} (g_R) \in[-0.021,0.021], & \operatorname{Im} (g_R) \in[-0.13,0.13] .
\end{array}
\end{align}
These results are summarized in Tab.~\ref{tab:results} together with the previous constraints from Ref. \cite{Deliot:2018jts}.

\begin{table}[h]
\centering
\small
\renewcommand{\arraystretch}{1.3}
\begin{tabular}{|c||c|c|c|c|c|c|}\hline
Limits & $ {\rm Re} ~V_R$& ${\rm Im} ~V_R$ & ${\rm Re} ~ g_L$& ${\rm Im}~ g_L$ & ${\rm Re}~ g_R$& ${\rm Im} ~g_R$ \\\hline \hline
Our method & [-0.037 , 0.037]& [-0.037 , 0.037] & [-0.028 , 0.028]& [-0.028 , 0.028] & [-0.021 , 0.021] & [-0.13 , 0.13]\\\hline
Previous & [-0.28 , 0.32]& [-0.30 , 0.30]& [-0.17 , 0.19]& [-0.19 , 0.18]& [-0.05 , 0.02]& [-0.11 , 0.10] \\\hline
\end{tabular}
\caption{The limited regions for the anomalous $Wtb$ couplings $V_R,g_L,g_R$ obtained from measurements of spin observables, together with previous constraints~\cite{Deliot:2018jts}.}
    \label{tab:results}
\end{table}

The projected constraints demonstrate that the complete $bW$ spin information provides strong sensitivity to anomalous $Wtb$ interactions for all parameters' analysis.
Compared with conventional measurements based mainly on the single top production cross sections and the $W$-boson helicity fractions, the present approach exploits additional information contained in the full spin-density matrix.

The substantial improvement in the sensitivity to $V_R$ and $g_L$ originates from the off-diagonal terms of the $bW$ spin density matrix. 
For these couplings, the interference with the SM contribution is suppressed by the bottom-quark mass in the diagonal elements, whereas unsuppressed linear contributions can remain in the off-diagonal components. 
Measurements based only on the $W$-boson helicity fractions therefore discard part of this interference information. 
By additionally including the $b$-quark polarization and the $bW$ spin correlations, our analysis recovers these linear contributions and consequently yields considerably stronger constraints on $V_R$ and $g_L$.

The behavior of $g_R$ is qualitatively different. 
Since the amplitude induced by $g_R$ can interfere directly with the dominant SM amplitude without bottom-mass suppression, its effects already appear linearly in the conventional $W$-polarization observables. 
The complete density-matrix analysis therefore provides complementary information but leads to a less pronounced relative improvement. 
This accounts for the enhanced sensitivity to $V_R$ and $g_L$, while the improvement for $g_R$ remains comparatively limited.

The sensitivity of the proposed method also depends on the preservation of the original $b$-quark polarization during hadronization into the $\Lambda_b$ baryon. This effect is characterized by the polarization-retention factor $r_i$, which directly controls the spin analyzing power of observables involving the $b$-quark spin. The benchmark limits above are obtained using the conservative value $r_i=0.5 $.
To quantify the impact of polarization retention, we repeat the analysis for
\begin{equation}
r_i\in[0.5,0.8].
\end{equation}

\begin{figure}[t]
    \centering
    \includegraphics[width=0.48\linewidth]{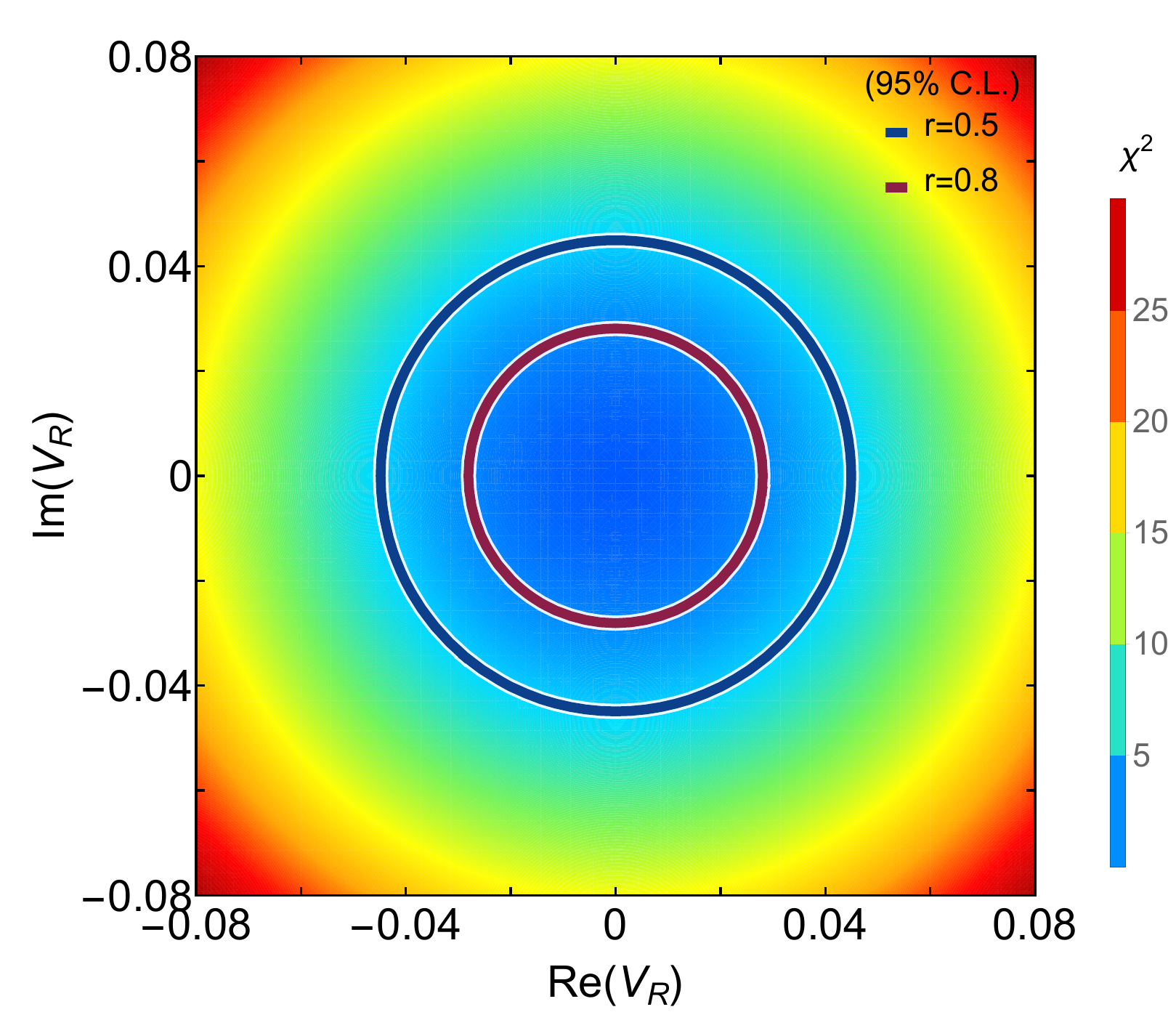}
    \includegraphics[width=0.48\linewidth]{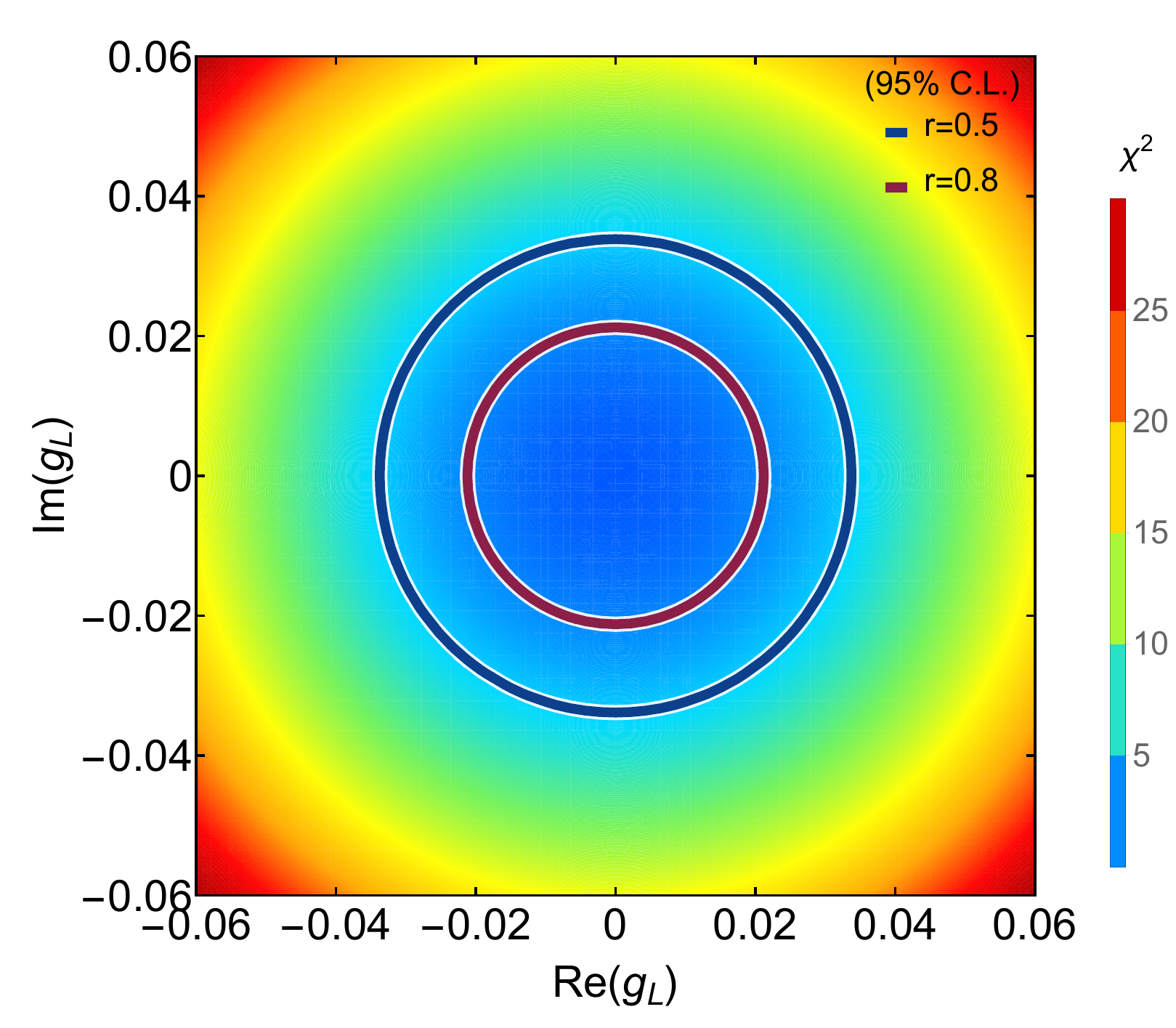}
    \vspace{0.3cm}
    \includegraphics[width=0.48\linewidth]{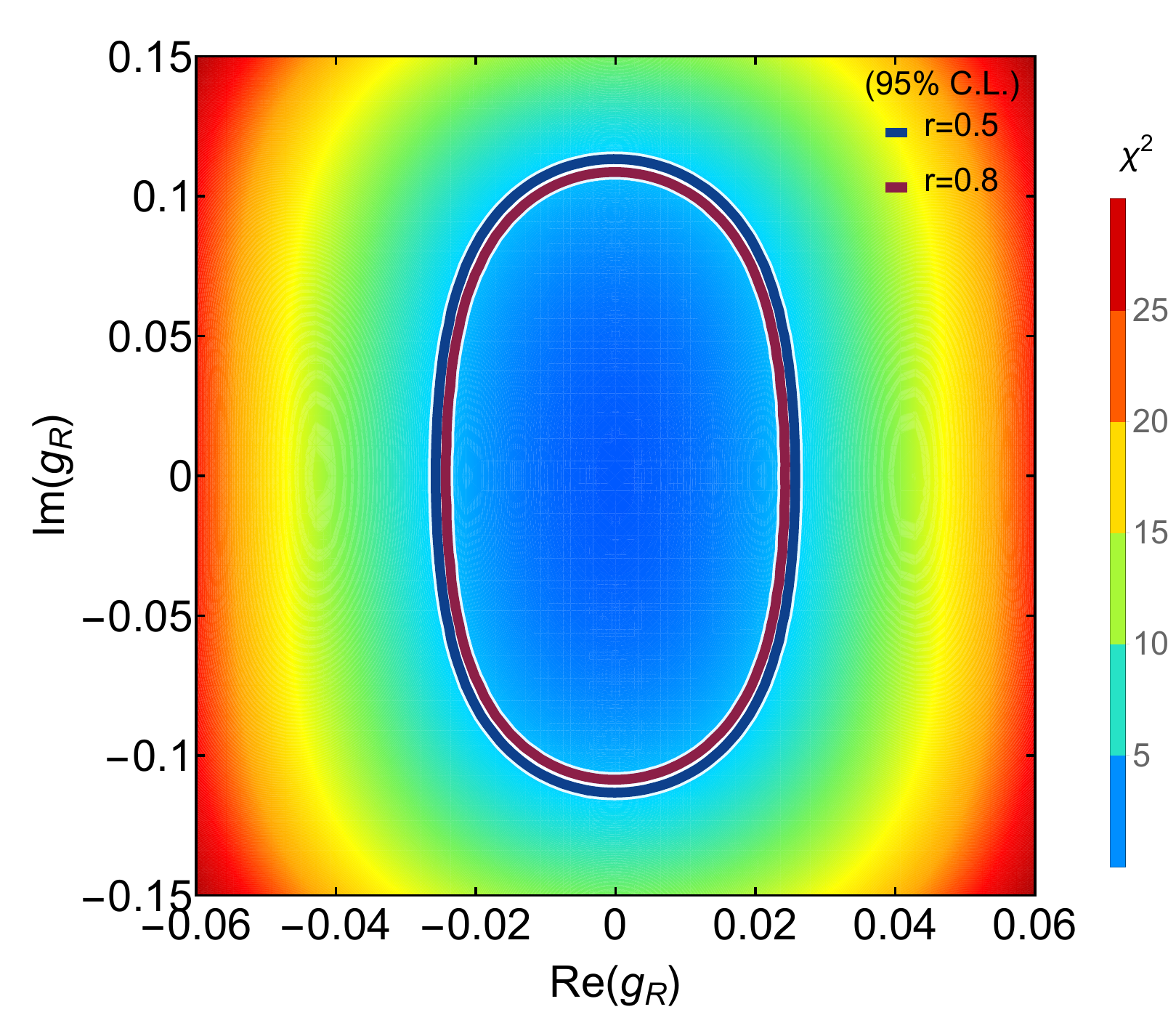}
    
    \caption{The expected statistical significance $\chi^2$ of new physics parameters $V_R,g_L,g_R$ as a function of retention factor $r$. The contour plots take $r\approx 0.5$ as a conservative estimation, and the colored lines give the 95\% confidence level($\Delta\chi^2=5.99$) for different $r\in [0.5,0.8]$.}
    \label{fig:chi2_contour}
\end{figure}

The resulting confidence contours in the anomalous coupling parameter space are shown in Fig.~\ref{fig:chi2_contour}. For two-dimensional parameter scans, the contours correspond to the 95\% confidence level,
\begin{equation}
\Delta\chi^2=5.99 .
\end{equation}
As $r_i$ increases, a larger fraction of the original $b$-quark polarization is retained by the $\Lambda_b$ baryon, thereby enhancing the sensitivity of observables involving the $b$-quark spin. Consequently, the allowed parameter regions for $V_R$ and $g_L$ become progressively smaller. However, the constraints on $g_R$ improve only modestly as $r_i$ increases, owing to the relatively weak sensitivity of these observables to this coupling.

Overall, the complete $b W$ spin-density matrix provides information complementary to conventional measurements based on the single top production cross sections and the $W$-boson helicity fractions. The largest gains are obtained for $V_R$ and $g_L$, for which off-diagonal spin correlations retain interference information that is suppressed in conventional observables. Improved $\Lambda_b$ reconstruction and a more precise determination of the $b \rightarrow \Lambda_b$ polarization-retention factors will therefore be important for realizing the full sensitivity of this method at the HL-LHC.

\section{Conclusion}
\label{sec:conclusion}

In this work, we have developed a spin-density-matrix framework for probing anomalous $Wtb$ interactions in top-quark pair production by exploiting the complete spin information of the $bW$ system. In addition to the conventional $W$-boson polarization observables, our approach incorporates the $b$-quark polarization and the spin correlations between the $b$ quark and the $W$ boson. These quantities can be accessed experimentally through the joint angular distributions of leptons from semileptonic $\Lambda_b$ and $W$-boson decays.

A key feature of the density-matrix description is that it retains the off-diagonal elements encoding spin coherence. For $V_R$ and $g_L$, the interference between the SM and anomalous amplitudes is suppressed by the bottom-quark mass in conventional observables such as the $W$-boson helicity fractions. In contrast, unsuppressed contributions linear in these couplings survive in the off-diagonal components of the $bW$ density matrix and become accessible once the joint $bW$ spin information is retained. This leads to substantially enhanced sensitivity to $V_R$ and $g_L$. The situation is different for $g_R$, which already interferes linearly with the dominant SM amplitude in conventional observables, so the additional spin information provides a smaller relative improvement in sensitivity.

We have evaluated the projected sensitivity at the HL-LHC with $\sqrt{s}=14~\mathrm{TeV}$ and an integrated luminosity of $3000~\mathrm{fb}^{-1}$. We considered semileptonic $t\bar t$ events in which the signal-side $W$ boson decays leptonically, the associated $b$ quark fragments into a semileptonically decaying $\Lambda_b$ baryon, and the opposite-side top quark decays hadronically. Incorporating the estimated signal efficiency and background dilution, and adopting $|\alpha_b|=|\alpha_W|=1$ and a conservative polarization-retention factor $r_i=0.5$, we obtain the following projected one-parameter 95\% confidence intervals:
\begin{align}
\begin{array}{ll}
\operatorname{Re} (V_R) \in[-0.037,0.037], & \operatorname{Im} (V_R) \in[-0.037,0.037], \\
\operatorname{Re} (g_L) \in[-0.028,0.028], & \operatorname{Im} (g_L) \in[-0.028,0.028],\\
\operatorname{Re} (g_R) \in[-0.021,0.021], & \operatorname{Im} (g_R) \in[-0.13,0.13] .
\end{array}
\end{align}
Compared with previous projections based primarily on the single top production cross sections and $W$-boson helicity fractions, the allowed ranges for both the real and imaginary parts of $V_R$ are reduced by approximately a factor of 8, while those for $g_L$ are reduced by nearly a factor of 6. For $g_R$, the additional spin information leads to a more modest improvement in $\operatorname{Re}g_R$, while the sensitivity to $\operatorname{Im}g_R$ remains comparable to that of the conventional approach. We also find that the constraints become stronger as the polarization-retention factor increases from $r_i=0.5$ to $0.8$, highlighting the importance of $b$-quark polarization retention during fragmentation and of a precise experimental determination of the corresponding retention factor. 

The present results should be interpreted as projections of the expected statistical sensitivity rather than as a full experimental analysis. A realistic measurement will require a detector-level treatment of the two-neutrino reconstruction ambiguity, angle-dependent efficiencies and resolutions, nontrivial background angular distributions, systematic uncertainties, and correlations among the extracted spin observables. In addition, simultaneous fits of multiple anomalous couplings will be important for accounting for possible correlations and degeneracies beyond the one-parameter scenarios considered here. Improved measurements of the $b\to\Lambda_b$ polarization-retention factors will also be essential for fully exploiting the $b$-quark spin information.

Our results demonstrate that the off-diagonal spin structure of the $bW$ system contains interference information that is not accessible through the $W$-boson helicity fractions alone. By retaining the $b$-quark polarization and its spin correlations with the $W$ boson, the proposed framework provides complementary sensitivity to the structure of the $Wtb$ interaction, particularly to anomalous couplings whose interference with the SM is suppressed in conventional observables. Measurements of the complete $bW$ spin structure therefore offer a promising avenue for precision tests of the $Wtb$ vertex at the HL-LHC.

\section{Acknowledgments}
The work of J.L. is supported by the National Science Foundation of China under Grant No.~12235001, No.~12475103 and State Key Laboratory of Nuclear Physics and Technology under Grant No. NPT2025ZX11.
The work of X.P.W. is supported by the National Science Foundation of China under Grant No. 12375095, and the Fundamental Research Funds for the Central Universities. 
The authors gratefully acknowledge the valuable discussions and insights provided by the members of the Collaboration of Precision Testing and New Physics.

\bibliographystyle{JHEP}
\bibliography{ref}

\newpage

\appendix

\end{document}